\documentclass[aps, 10pt,  twocolumn]{revtex4-2}
\usepackage{graphicx, epsfig}
\usepackage[english]{babel} 
\usepackage{bm}
\usepackage{comment}
\usepackage{amsmath}     
\usepackage{amsfonts}
\usepackage{hyperref}
\usepackage{xcolor}
\usepackage{soul}

\begin{document}

\preprint{APS/123-QED}

\title{Far-field heat transfer and monochromatic thermal currents in a cylindrical \\ nonreciprocal cavity}

\author{Guillem Masdemont}
\author{Julien Legendre}%
\email{Julien.legendre@icfo.eu} 
\author{Georgia T. Papadakis}%
\email{Georgia.Papadakis@icfo.eu}
\affiliation{%
 ICFO-Institut de Ciències Fotòniques, The Barcelona Institute of Science and Technology, 08860 Castelldefels (Barcelona), Spain
}%

\date{\today}
\begin{abstract}

Breaking Kirchhoff’s law of thermal radiation yields new opportunities in one-way radiative thermal transport and circuitry. We investigate its consequences in the far-field regime in cylindrical cavities, by employing a specular ray-tracing algorithm. At thermal equilibrium, we show that violation of Kirchhoff’s law yields non-vanishing heat rectification coefficients within different sections of the cavity, which can be tuned for perfect rectification and circulation, while internal monochromatic currents vanish due to the intrinsic coupling between emission and absorption at specular surfaces. This constraint is lifted under nonequilibrium conditions, where rotational heat fluxes within the cavity can be precisely controlled by appropriately combining reciprocal and nonreciprocal materials. These findings open new avenues for thermal management and provide design principles for nonreciprocal photonic devices.
\end{abstract}

\maketitle

\section{\label{Introduction} Introduction}
The classical form of Kirchhoff’s law of thermal radiation~\cite{KirchhoffUeberVerhaltnisszwischenEmissionsvermogenundAbsorptionsvermogenKorperfurWarmeundLicht1860} states that, at a given frequency $\omega$, monochromatic radiation absorbed from a polar–azimuthal direction \( (\theta, \varphi) \) must be re-emitted equally in that same direction:
\begin{equation}\label{eq:kirchhoff_law}
    \epsilon(\omega, \theta, \varphi) = \alpha(\omega, \theta, \varphi).
\end{equation}

Recent works have shown that Kirchhoff’s law can be violated along certain directions \cite{zhuNearcompleteViolationDetailed2014,
ZhangValidityKirchhoffslawsemitransparentfilmsmadeanisotropicmaterials2020,parkViolatingKirchhoffsLaw2021, fernandesEnhancingDirectionalViolation2023}. This effect finds practical applications in renewable energy, circuitry and communications
 ~\cite{YangNonreciprocalthermalphotonics2024, zhangNonreciprocalThermalPhotonics2022}. Within the context of radiative heat engines, nonreciprocal thermal emission enables dramatic performance enhancement \cite{puschFundamentalEfficiencyBounds2019,
giteauThermodynamicPerformanceBounds2023,giteauFundamentalLimitationsThermoradiative2025} due to improved light-harvesting. For example, breaking reciprocity between absorption and thermal emission improves performance in photovoltaics~\cite{GreenTimeAsymmetricPhotovoltaics2012, jafarighalekohnehNonreciprocalSolarThermophotovoltaics2022, ParkReachingUltimateEfficiencySolarEnergyHarvestingNonreciprocalMultijunctionSolarCell2022}, thermophotovoltaics ~\cite{
picardiNonreciprocityTransmissionMode2024, OmairUltraefficientthermophotovoltaicpowerconversionbandedgespectralfiltering, 
NefzaouiSelectiveemittersdesignoptimizationthermophotovoltaicapplications2012}, and radiative cooling~\cite{fangRadiativeCoolingVertical2024} via redirecting emitted radiation towards directions and channels where it can be maximally utilized. Beyond light-harvesting, breaking Kirchhoff’s law becomes relevant in thermal circuitry and potentially information technology, via effects such as the the photon thermal Hall effect~\cite{Ben-AbdallahInverseSpinThermalHallEffectNonreciprocalPhotonicSystems2025,GuoRelationphotonthermalHalleffectpersistentheatcurrentnonreciprocalradiativeheattransfer2019, ottAnomalousPhotonThermal2020}, gyrotropic heat engines~\cite{guoSingleGyrotropicParticle2021}, and persistent equilibrium heat currents in many-body systems~\cite{zhuPersistentDirectionalCurrent2016, zhuTheoryManybodyRadiative2018, biehsPersistentEnergyCurrents2025}. 

Breaking reciprocity between thermal emission and absorption can also lead to nonreciprocal radiative heat transfer \cite{fanNonreciprocalRadiativeHeat2020, zhuTheoryManybodyRadiative2018, GuoRelationphotonthermalHalleffectpersistentheatcurrentnonreciprocalradiativeheattransfer2019}. Several material systems have been explored, theoretically, as platforms for nonreciprocal radiative heat flow, ranging from magneto-optical materials~\cite{ abrahamekerothAnisotropicThermalMagnetoresistance2018, ZhaoNearcompleteviolationKirchhoffslawthermalradiation03magneticfield2019, ShayeganNonreciprocalinfraredabsorptionresonantmagnetoopticalcouplingInAs2022} to Weyl semimetals~\cite{zhaoAxionFieldEnabledNonreciprocalThermal2020, tsurimakiLargeNonreciprocalAbsorption2020, guoRadiativeThermalRouter2020}, and metasurface-based designs that leverage multiple diffraction channels and predict near-complete violation of Kirchhoff’s law ~\cite{ZhaoNonreciprocalThermalEmittersUsingMetasurfacesMultipleDiffractionChannels2021}, and recent experimental results ~\cite{ShayeganDirectobservationviolationKirchhoffslawthermalradiation2023, ZhaoRadiationImbalanceNewMaterialEmitsBetterItAbsorbs2025, ZhangObservationStrongNonreciprocalThermalEmission2025} confirm the suitability of these systems for breaking reciprocity in thermal radiation. Measurements have confirmed pronounced differences between emission and absorption, per direction, with several works reporting broadband and high-temperature demonstrations of nonreciprocal thermal emission~\cite{ShayeganBroadbandnonreciprocalthermalemissivityabsorptivity2024,nabaviHighTemperatureStrongNonreciprocal2025}. In order to maximize this nonreciprocal effect, multilayer architectures and optimization-based emitter designs are being considered ~\cite{wangMaximalViolationKirchhoffs2023,DoAutomateddesignnonreciprocalthermalemittersBayesianoptimization2025}.

In parallel to progress in uncovering nonreciprocal thermal effects, considerable effort has been devoted to revisiting the theoretical foundations of Kirchhoff’s law. Its classical formulation is fundamentally grounded in Lorentz reciprocity~\cite{kongElectromagneticWaveTheory1986} and the linear constitutive relations governing material response~\cite{MackayTransferMatrixMethodElectromagneticsOpticsSpringerLink2020}. Recent studies have introduced a symmetry-based classification of nonreciprocal thermal emitters and absorbers ~\cite{MillerUniversalmodalradiationlawsallthermalemitters2017, khandekarNewSpinresolvedThermal2020} and revealed that, under certain conditions, directional emission and absorption can be tuned independently~\cite{GuoAdjointKirchhoffsLawGeneralSymmetryImplicationsAllThermalEmitters2022}. Mathematically, this distinction originates from the structure of the constitutive relations and asymmetries of the dielectric and magnetic response tensors~\cite{KongTheoremsbianisotropicmedia1972, GuoAdjointKirchhoffsLawGeneralSymmetryImplicationsAllThermalEmitters2022,NovotnyPrinciplesNanoOptics2012}. Materials that allow fully independent control of directional emission and absorption must support multiple reflection channels, typically through diffraction \cite{ZhaoNonreciprocalThermalEmittersUsingMetasurfacesMultipleDiffractionChannels2021}. In contrast, bodies displaying specular reflection, whether reciprocal or not, obey a generalized form of Kirchhoff’s law: emissivity in a given direction is equal to absorptivity in the opposite one~\cite{YangPolarimetricanalysisthermalemissionbothreciprocalnonreciprocalmaterialsusingfluctuationelectrodynamics2022, ZhaoNonreciprocalThermalEmittersUsingMetasurfacesMultipleDiffractionChannels2021}
\begin{equation}\label{eq:specular_emission}
    \epsilon(\omega, \theta, \varphi) = \alpha(\omega, \theta, \varphi + \pi).
\end{equation}
 When the permittivity and magnetic permeability tensors are symmetric, specifically, when the material is invariant under the adjoint transformation~\cite{GuoAdjointKirchhoffsLawGeneralSymmetryImplicationsAllThermalEmitters2022}, the emissivity and absorptivity profiles become symmetric with respect to the normal direction. In this case, the generalized law reduces to the classical form of Kirchhoff’s law in Eq.~\eqref{eq:kirchhoff_law}. 

So far, studies investigating nonreciprocal radiative heat transfer and radiative thermal conductance remain mostly limited to planar and near-field geometries~\cite{zhuPersistentDirectionalCurrent2016, zhuTheoryManybodyRadiative2018,fanNonreciprocalRadiativeHeat2020}, with the exception of the recent work in ~\cite{7p58-n6yv}, which considers a two-dimensional, triangular geometry. The implications of nonreciprocal radiative heat flow in nonplanar and more complex configurations remain less understood. 

In this work, we investigate how nonreciprocity manifests in far-field radiative heat transfer in a hollow, infinitely long cylindrical cavity. This geometry is chosen as a simple and suitable platform for hosting persistent thermal currents, however it resists analytical treatment. To make the problem tractable, we discretize the cavity walls into vertically aligned, identical elements composed of either reciprocal or nonreciprocal materials. Using a custom-made ray-tracing algorithm tailored to this geometry, we demonstrate that nonreciprocity yields nonvanishing heat rectification coefficients, \textit{even at thermal equilibrium}. Despite these nonvanishing heat rectification coefficients, owning to the generalized form of Kirchhoff's law in Eq.~\eqref{eq:specular_emission}, persistent thermal currents do not arise within the cavity. Under nonequilibrium conditions, by contrast, nonreciprocity enables controllable, rotational heat fluxes, providing pathways for directional thermal transport in the far field.

The paper is structured as follows: Sec.~\ref{sec: Mathematical Background} introduces the mathematical formalism. Sec.~\ref{Sec: Numerical simulations} presents the numerical simulations and validates the theoretical constraints. We conclude in Sec.~\ref{Sec: Conclusions}. 
 
\section{Formalism}\label{sec: Mathematical Background}

\begin{figure}[t]
    \centering \includegraphics[width=0.8\linewidth]{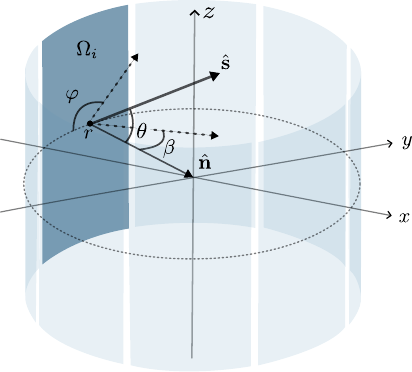}
    \caption{ Representation of the geometry under consideration. The cylindrical surface is partitioned into $n = 8$ evenly spaced vertical elements, assumed infinite along the $z$-direction. Each element $\Omega_i$, for $i=1, \dots, n$, is defined in Eq.~\eqref{eq cylinder partition} and assigned either a reciprocal or nonreciprocal material exhibiting specular reflection. At each surface point $\mathbf{r}$, emission and absorption obey the generalized Kirchhoff's law in Eq. \eqref{eq:specular_emission}, and are specified for every direction $\hat{\mathbf{s}}$ by the polar and azimuthal angles \((\theta, \varphi)\), measured relative to the local surface normal $\hat{\mathbf{n}}$. The horizontal angle \( \beta \in (-\pi/2, \pi/2) \) characterizes the projection of \( \hat{\mathbf{s}} \) on to the xy-plane with respect to \( \hat{\mathbf{n}} \).}       
    \label{fig: cylinder-emission}
\end{figure}

Throughout this work, we consider a cylindrical cavity that is infinite along the $z$-direction and immersed in vacuum (Fig. \ref{fig: cylinder-emission}). Its surface is defined by 
\begin{equation*}
(\alpha, z) \in [0, 2\pi) \times \mathbb{R} 
\xrightarrow{\psi} 
(R \cos\alpha, R \sin\alpha, z) \in \mathbb{R}^3,
\end{equation*}
where the radius $R > 0$ is fixed and assumed to be much larger than the thermal wavelength, thereby ensuring operation in the far-field regime. Our analysis is restricted to the cavity by assuming the surface to be opaque. 

The cylindrical surface is partitioned into $n$ adjacent elements $\Omega_i$ of identical geometry, defined as
\begin{equation}\label{eq cylinder partition}
		\Omega_i = \psi \left(\left[\frac{2 \pi}{n} (i-1) , \frac{2 \pi}{n} i \right), \mathbb{R} \right), \quad i \in \{1, \dots, n\}.
\end{equation}

Each element \(\Omega_i\) is assigned either a reciprocal or nonreciprocal material, specified through its permittivity tensor. As indicated in Fig. ~\ref{fig: cylinder-emission}, at each point on the cylinder's surface, the thermal emissivity is defined with respect to the local azimuthal-polar coordinates \((\theta, \varphi)\), relative to the normal vector $\hat{\mathbf{n}}$. At thermal equilibrium, no conduction occurs between adjacent elements, since the absence of temperature gradients precludes any net heat flow according to Fourier's law.

We develop a specular ray-tracing algorithm tailored to the cylindrical cavity, constrained by the generalized Kirchhoff's law in Eq.~\eqref{eq:specular_emission}. This approach is well-suited to materials exhibiting specular reflection \cite{YangPolarimetricanalysisthermalemissionbothreciprocalnonreciprocalmaterialsusingfluctuationelectrodynamics2022}. 
Moreover, although standard ray-tracing techniques can incorporate interference effects such as Fabry–Pérot resonances, these contributions are negligible in the present context owing to the incoherent character of thermal radiation and the large separation between surfaces with respect to the thermal wavelength.

This study employs two primary metrics, introduced in Sections~\ref{sec: Transmission coeficient} and~\ref{sec: Heat flux vector field}: the monochromatic transmission coefficient $S_{i \to j}(\omega)$ between elements $\Omega_i$ and $\Omega_j$ of the cylindrical cavity, expressing the normalized electromagnetic energy received by element $\Omega_j$ from element $\Omega_i$ at frequency $\omega$, and the heat-flux vector field $\mathbf{v}$, expressing the net thermal flow obtained (at each point) by integrating the radiative contributions from all directions. Both are defined inside the cavity and along its boundary. Without loss of generality, the analysis is restricted to monochromatic radiation.

\subsection{Monochromatic transmission coefficient}\label{sec: Transmission coeficient}

The average heat absorbed by an element \( \Omega_i \), denoted by \( Q_i \), is computed as the time-ensemble average of the Poynting vector flux through its surface \cite{zhangNanoMicroscaleHeat2020, BergmanFundamentalsHeatMassTransfer2011}:  
\begin{equation*}\label{eq:Qdot}
    Q_i = \int_{\Omega_i} \left\langle \mathbf{S}(\mathbf{r}, t) \cdot \hat{\mathbf{n}} \right\rangle \, d\Omega_i,
\end{equation*}
where \( \mathbf{r} \in \Omega_i \), \( \mathbf{S}(\mathbf{r}, t) \) is the instantaneous Poynting vector at point \( \mathbf{r} \) and time \( t \), \( \hat{\mathbf{n}}(\mathbf{r}) \) is the inward unit normal at \( \mathbf{r} \), and \( \langle \cdot \rangle \) denotes the time-ensemble average. Through the fluctuation-dissipation theorem \cite{NovotnyPrinciplesNanoOptics2012}, the integrand takes the form
\begin{equation}\label{eq: heat flux - transmission coefficient}
\begin{aligned}
\langle \mathbf{S}(\mathbf{r}, t) \cdot \hat{\mathbf{n}} \rangle
=  \int_0^\infty \Theta(\omega, T) \, \frac{d\omega}{2\pi} 
\int \frac{d\mathbf{k}_{\parallel}}{(2\pi)^2}  \, S(\mathbf{r}, \mathbf{k}_{\parallel}, \omega)
\end{aligned}
\end{equation}
where $\Theta(\omega, T) = \hbar \omega \left[\frac{1}{2} + \frac{1}{\exp(\hbar \omega / k_\text{B} T) - 1} \right]$ denotes the mean photon energy at frequency $\omega$ and temperature $T$, with $\hbar$ and $k_\text{B}$ representing the reduced Planck constant and the Boltzmann constant, respectively. The function $S(\mathbf{r}, \mathbf{k}_{\parallel}, \omega)$ is henceforth termed directional monochromatic transmission coefficient at position $\mathbf{r}$ and frequency $\omega$, for propagation direction defined by the local angles $(\theta, \varphi)$, encoded in the in-plane wavevector $\mathbf{k}_{\parallel}$. 

The absorbed heat at a point $\mathbf{r} \in \Omega_i$ can be expressed as the sum of contributions from all other elements $\Omega_j$ by decomposing the transmission coefficient into element-specific contributions:
\begin{equation}\label{eq: S decomposition}
    S(\mathbf{r}, \mathbf{k}_{\parallel}, \omega) =  \sum_{j = 1}^{n} S_j(\mathbf{r}, \mathbf{k}_{\parallel}, \omega),
\end{equation}
where $S_j(\mathbf{r}, \mathbf{k}_{\parallel}, \omega)$ represents the directional transmission coefficient at point $\mathbf{r}$ due to element $\Omega_j$. The corresponding spectral transmission coefficient at $\mathbf{r}$, associated with $\Omega_j$, is defined as

\begin{equation*}
S_j(\mathbf{r}, \omega) = \frac{1}{\pi} \left( \frac{c}{\omega} \right)^2 \sum_{\sigma \in \{s, p\}} \int d\mathbf{k}_{\parallel} \, S_j(\mathbf{r}, \mathbf{k}_{\parallel}, \omega),
\end{equation*}
where the prefactor ensures normalization with respect to blackbody transmission~\cite{planckTheoryHeatRadiation1991}, while $s$ and $p$ express the two orthogonal linear polarizations. Exploiting translational invariance along the $z$-axis, the transmission coefficient from $\Omega_j$ to $\Omega_i$, denoted $S_{j \to i}(\omega)$, is obtained by averaging $S_j(\mathbf{r}, \omega)$ over the arc corresponding to $\Omega_i$:
\begin{equation}\label{eq:Sij}
S_{j \to i}(\omega) = \frac{n}{2\pi} \int_{\frac{2\pi R}{n}i}^{\frac{2\pi R}{n}(i+1)} S_j(\mathbf{r}, \omega) \, d\alpha.
\end{equation}

The resulting heat rectification coefficient between elements $\Omega_i$ and $\Omega_j$, denoted $H_{i j}(\omega)$, is defined as the directional imbalance of their mutual exchange,
\begin{equation*}
H_{i  j}(\omega) = S_{i \to j}(\omega) - S_{j \to i}(\omega).
\end{equation*}
By construction, $H_{ij}(\omega) = -H_{ji}(\omega)$, and can therefore take both positive and negative values. A positive value indicates a net monochromatic heat flux from element $i$ to element $j$, while a negative value corresponds to a monochromatic net flux in the opposite direction. 

In two-body systems, such directional asymmetries are quantified via the heat rectification ratio \cite{7p58-n6yv}. However, this pairwise metric is not directly applicable to a many-body configurations. To alleviate this restriction, we introduce the nonreciprocity factor \(\zeta\), defined as
\begin{equation}\label{Eq:non-reciprocity-factor}
    \zeta =  \frac{1}{n (n-1)} \sum_{i \neq j}^n \gamma_{ij}.
\end{equation}
Here, $\gamma_{ij} = |S_{i \to j} - S_{j \to i}| / (S_{i \to j} + S_{j \to i})$. The quantity $\zeta$ can thus be viewed as a generalization of the heat rectification ratio to a many body configuration, accounting for all pairwise contributions. For reciprocal configurations, \(\zeta = 0\), while nonreciprocal materials yield \( 0 < \zeta \leq  1\) owing to the presence of nonvanishing terms in \(\gamma_{ij}\).

\subsection{Monochromatic heat flux vector field}
\label{sec: Heat flux vector field}

Let $\overline{\mathcal{C}}$ denote the hollow cylindrical region defined as
$$\overline{\mathcal{C}} = \left\{ (r \cos \theta, r \sin \theta, z) \,\middle|\, 0 \leq r \leq R,\; 0 \leq \theta < 2\pi,\; z \in \mathbb{R} \right\},$$
with boundary $$\partial \mathcal{C} = \left\{ (R \cos \theta, R \sin \theta, z) \,\middle|\,  0 \leq \theta < 2\pi,\; z \in \mathbb{R} \right\}$$ and interior $\mathcal{C} = \overline{\mathcal{C}} \setminus \partial \mathcal{C}$. The monochromatic heat flux vector at point $\mathbf{p} \in \overline{\mathcal{C}}$, denoted $\mathbf{v}(\mathbf{p}, \omega)$, is defined as the directional average of the spectral heat flux over all directions of the unit sphere. In particular, for points inside the cavity ($\mathbf{p} \in \mathcal{C}$), emission and absorption at $\mathbf{p}$ are neglected, so the flux reduces to the contribution from incoming radiation
\begin{equation}
\label{eq:flux-interior}
\mathbf{v}(\mathbf{p}, \omega) = \frac{1}{\pi} \int_{\mathbb{S}^2} S(\mathbf{p}, \hat{\mathbf{s}}, \omega) \, \hat{\mathbf{s}} \, d\hat{\mathbf{s}},
\end{equation}

whereas for points on the boundary ($\mathbf{p} \in \partial \mathcal{C}$), the flux accounts for both directional emission and incident radiation
\begin{equation}
\label{eq:flux-boundary}
\mathbf{v}(\mathbf{p}, \omega) = \frac{1}{\pi} \int_{\mathbb{S}_+^{2}} \left[ \epsilon(\mathbf{p}, \hat{\mathbf{s}}, \omega) - S(\mathbf{p}, -\hat{\mathbf{s}}, \omega) \right] \hat{\mathbf{s}} \, d\hat{\mathbf{s}}.
\end{equation}

Here, $\mathbb{S}^2$ is the unit sphere in $\mathbb{R}^3$, and $\mathbb{S}_+^2 = \{ \hat{\mathbf{s}} \in \mathbb{S}^2 \mid \hat{\mathbf{s}} \cdot \hat{\mathbf{n}} > 0 \}$ denotes the inward-facing hemisphere at a boundary point $\mathbf{p}$, defined with respect to the local inward unit normal $\hat{\mathbf{n}}$. The direction vector $\hat{\mathbf{s}}$ is parameterized by spherical angles $(\theta, \varphi)$, and the differential solid angle element is given by $d\hat{\mathbf{s}} = \sin\theta \, d\theta \, d\varphi$. The function $S(\mathbf{p}, \hat{\mathbf{s}}, \omega)$ denotes the monochromatic directional transmission coefficient at point $\mathbf{p}$ in direction $\hat{\mathbf{s}}$, while $\epsilon(\mathbf{p}, \hat{\mathbf{s}}, \omega)$ denotes the corresponding directional emissivity, defined only on the boundary. The factor $\pi$ ensures normalization with respect to blackbody emission.  

In Eq.~\eqref{eq:flux-interior}, $S(\mathbf{p}, \hat{\mathbf{s}}, \omega)$ includes only incoming radiative contributions from the surfaces, whereas Eq.~\eqref{eq:flux-boundary} also incorporates absorption at point $\mathbf{p}$. Further details of the computations are provided in Appendix~\ref{Sec: computation S appendix}, and numerical simulations of these quantities are presented in the following section.

\section{Results \& Discussion}\label{Sec: Numerical simulations}

We present numerical simulations of the monochromatic transmission coefficient and the corresponding monochromatic heat-flux vector field for the the cylindrical configuration introduced in Section \ref{sec: Mathematical Background}. We prescribe an emissivity profile $ \epsilon(\theta, \varphi)$ for each element of the configuration, according to the azimuthal–polar angle convention illustrated in Fig.~\ref{fig: cylinder-emission}, and restrict our analysis to specularly reflecting media that satisfy the generalized Kirchhoff’s law given in Eq.~\eqref{eq:specular_emission}. As a benchmark, we consider reciprocal blackbody emission with uniform emissivity, $ \epsilon(\theta, \varphi) = 1 $. This blackbody case is compared with an ideal nonreciprocal scenario based on which radiation is predominantly absorbed when incident from the left hemisphere and emitted preferentially into the right hemisphere, termed \textit{on-off} henceforth.

Mathematically, this behavior is modeled by the smooth, angle-dependent emissivity profile $\epsilon(\theta, \varphi) = \frac{1}{2} + \frac{\arctan(k \beta)}{\pi}$, where $k \gg 1$ and $\beta \in (-\pi/2, \pi/2)$ is a function of the spherical coordinates $(\theta, \varphi)$  representing the azimuthal angle of the emission direction $\hat{\mathbf{s}}$ projected onto the $xy$-plane, as illustrated in Fig.~\ref{fig: cylinder-emission}. In this parametrization, $\epsilon \lesssim 1$ as $\beta \to \pi/2$, and $\epsilon \gtrsim 0$ as $\beta \to -\pi/2$. This describes a strong directional asymmetry and maximal violation of Kirchhoff’s law in its classical form. For completeness, Appendix~\ref{sec: weyl semimetal} presents results for a realistic Weyl semimetal, which qualitatively reproduce the behavior of the \textit{on-off} material, but exhibit reduced directional selectivity in radiative exchange.

\begin{figure}[]
    \centering
    \includegraphics[width=0.7\linewidth]{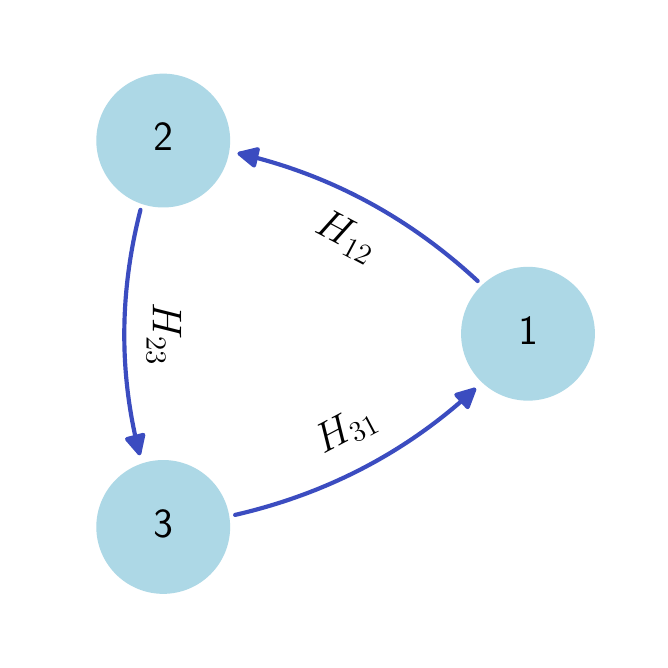}
    \caption{Schematic representation of a directed graph illustrating the monochromatic heat exchange in a three element partition of the cylinder at thermal equilibrium. Each node $i$ corresponds to an element $\Omega_i$, while arrows indicate the directional imbalance in the heat rectification coefficient \( H_{ij} \). For a reciprocal material configuration, symmetry in the emissivity profile with respect to \(\theta\) ensures \( S_{i \to j} = S_{j \to i} \), leading to \( H_{ij} = 0 \), and thus balanced radiative exchange. By contrast, nonreciprocal materials with asymmetric emissivity profiles yield \( S_{i \to j} \neq S_{j \to i} \), resulting in \( H_{ij} \neq 0 \), i.e, a nonzero heat rectification coefficient. For the idealized \textit{on-off} material, \( |H_{ij}| = 0.26 \), whereas for a Weyl semimetal it is reduced to \( |H_{ij}| = 0.028\) (see Appendix~\ref{sec: weyl semimetal}).}
    \label{fig:3 element configuration - WS}
\end{figure}

We begin by analyzing a configuration at thermal equilibrium, in which the cylinder is partitioned into three elements composed of the same material. To visualize the results, we employ the directed graph representation shown in Fig.~\ref{fig:3 element configuration - WS}, where each node corresponds to an individual element and the arrows denote the heat rectification coefficient between element pairs.

The heat rectification coefficient $H_{ij}$ quantifies the directional imbalance in monochromatic radiative exchange between elements $i$ and $j$. For reciprocal materials, the symmetry of the emissivity profile with respect to the polar angle $\theta$ guarantees $S_{i \to j} = S_{j \to i}$, yielding $H_{ij} = 0$. In contrast, for nonreciprocal materials, this symmetry is broken, hence \( S_{i \to j} \neq S_{j \to i} \), and consequently \( H_{ij} \neq 0 \). Owing to the \( C^3 \) rotational symmetry of the system, such imbalance gives rise to closed-loop heat-exchange channels that exhibit net circulation, with the orientation set by the asymmetric emissivity profile. This phenomenon is analogous to the cyclic transmission coefficient reported in a three-body configuration in the near-field in ~\cite{zhuPersistentDirectionalCurrent2016}, and in a triangular geometry in the far-field in~\cite{7p58-n6yv}. Our results remain consistent with the second law of thermodynamics, as the heat rectification coefficients satisfy $H_{1  2} = H_{2  3} = H_{3  1}$ as expected in equilibrium. Quantitatively, for the \textit{on–off} material, we obtain $|H_{ij}| = 0.26$. 

\begin{figure*}[t]
    \centering
    \includegraphics[width=1\linewidth]{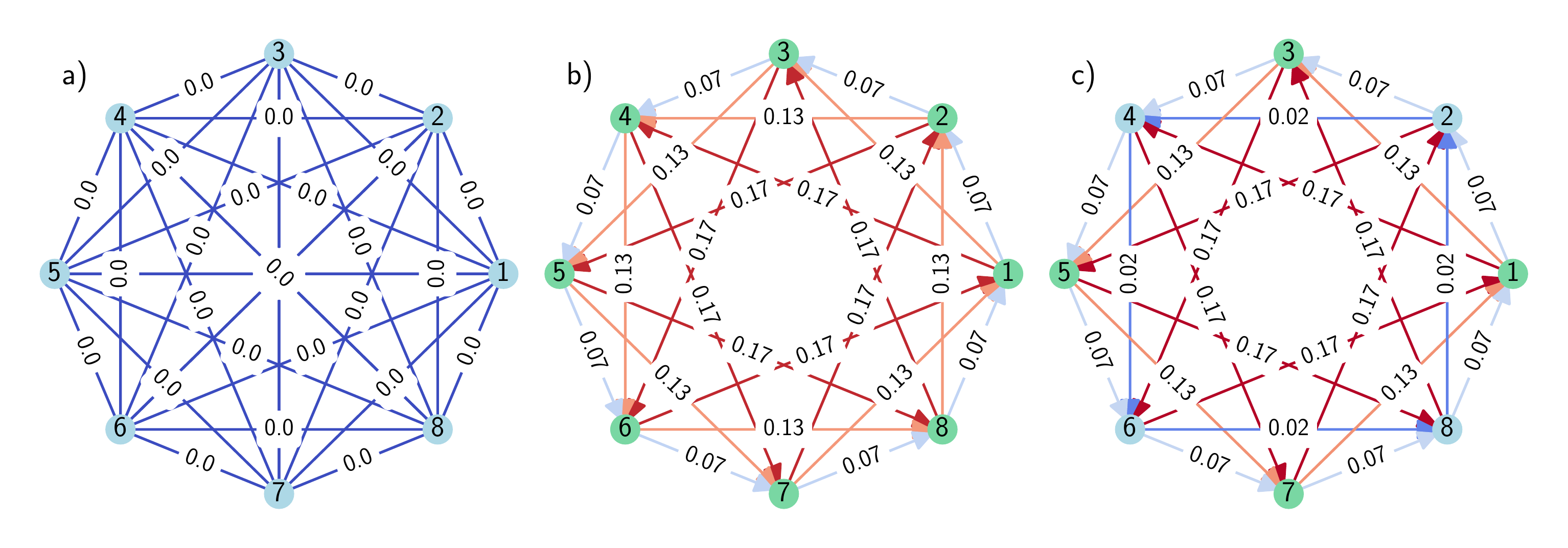}
    \caption{Graph representations of the heat rectification coefficient, defined as \( H_{i  j} = S_{i \to j} - S_{j \to i} \), for a cylindrical configuration partitioned into eight elements. Arrows indicate the direction and magnitude of net radiative exchange between element pairs; absence of an arrow implies \( H_{i  j} = 0 \). (a) A reciprocal material with a symmetric emission and absorption profile is considered. Due to reciprocity and angular symmetry, the heat rectification coefficients vanish for all element pairs, i.e., \( H_{i  j} = 0 \), consistent with detailed balance and reciprocity constraints. (b) A nonreciprocal idealized \textit{on–off} emitter is used, characterized by high emissivity into the right hemisphere ($\beta>0$) and high absorptivity from the left ($\beta<0$). This asymmetry leads to a pronounced directional transmission pattern, with significant nonzero values of \( H_{i  j} \), forming circular heat channels. The configuration exhibits \( C_8 \) rotational symmetry and enhances directional heat rectification coefficients by nearly an order of magnitude compared to realistic nonreciprocal materials. (c) A mixed configuration where odd-numbered elements in green are nonreciprocal (\textit{on–off} type), and even-numbered elements in blue are reciprocal blackbodies. Despite the presence of only four nonreciprocal elements, nonzero heat rectification coefficients emerge across nearly all element pairs due to indirect radiative coupling mediated by reflection and emission. In all cases, the total net power exchange at each element vanishes, as required by the second law of thermodynamics. 
}
    \label{fig: Transmission coefficient graphs}
\end{figure*}

Next, we extend the analysis to a configuration in which the cylinder is partitioned into eight elements, considering three distinct scenarios: a fully reciprocal configuration, a fully nonreciprocal configuration, and a mixed configuration combining reciprocal and nonreciprocal elements. The corresponding results are presented in Figure~\ref{fig: Transmission coefficient graphs}. Panel (a) shows the directed graph for a fully reciprocal blackbody configuration. As in the three-element case of Fig.~\ref{fig:3 element configuration - WS}, reciprocity enforces a symmetric emissivity profile. Consequently, the radiative exchange between any pair of elements is balanced, i.e., \( S_{i \to j} = S_{j \to i} \), and the heat rectification coefficient vanishes, \( H_{ij} = 0 \). Simulations performed for other reciprocal materials yield qualitatively identical results.

Panels (b) and (c) correspond to configurations involving nonreciprocal materials, which exhibit asymmetric emissivity profiles. In panel (b), the cylinder is entirely composed of the \textit{on–off} material. Owing to the $C_8$ rotational symmetry of the system, the resulting directed graph inherits rotational invariance. The nonvanishing heat rectification coefficients ($H_{ij} \neq 0 $) reveals the emergence of closed heat exchange channels in nonplanar geometries, which appear as circulation patterns in the directed graph and stem from the asymmetric emissivity profile. The strongest transmission occurs in directions close to the surface normal, in accordance with Lambert’s cosine law. 
The \textit{on–off} material produces the highest heat rectification coefficients, exceeding those of realistic Weyl semimetals by approximately an order of magnitude as detailed in Appendix~\ref{sec: weyl semimetal}.

In panel (c), we examine a configuration in which reciprocal and nonreciprocal elements alternate: elements with odd indices (green nodes) are assigned the \textit{on–off} nonreciprocal material, while even-indexed elements (blue nodes) are modeled as reciprocal blackbodies. The results exhibit nonzero heat rectification coefficient not only between nonreciprocal elements, but also between reciprocal ones, although the latter remain significantly weaker as compared to the fully nonreciprocal case shown in panel (b). This occurs because, while direct exchange between reciprocal elements vanishes due to reciprocity, as enforced by the ray-tracing algorithm, indirect coupling through reflections involving nonreciprocal elements generally yields \( H_{ij} \neq 0 \). Consequently, even configurations containing a single nonreciprocal element embedded in a reciprocal background can exhibit nonvanishing heat rectification coefficients across the entire system.

The deviation from reciprocity in each configuration is quantified by the nonreciprocity factor $\zeta_8$, as defined in Eq.~\eqref{Eq:non-reciprocity-factor}. For the reciprocal configuration in panel (a), we obtain $\zeta_8 = 0$, as expected. The fully nonreciprocal \textit{on–off} configuration in panel (b) yields $\zeta_8 = 0.85$, which asymptotically approaches the theoretical maximum $\zeta_\infty = 1$ in the limit of infinitely many elements ($n \to \infty$). In this case, $\gamma_{ij} = 1$ for all pairs of elements, hence perfect rectification is achieved throughout the cylinder. This is in contrast to the triangular geometry reported in~\cite{7p58-n6yv}, which requires an angularly selective emissivity profile, acting as a perfect mirror at specific angles while exhibiting the \textit{on-off} behavior that we introduced above at others. The mixed configuration in panel (c), consisting of alternating reciprocal and nonreciprocal elements, gives $\zeta_8 = 0.72$. This demonstrates that even with fewer nonreciprocal elements, substantial directional imbalance in the heat flow can be induced. 

Our findings are consistent with the second law of thermodynamics, since the sum of heat rectification coefficients entering and leaving each element vanishes in all cases considered. This consistency is further validated by symmetry considerations and local energy conservation at every surface point.

In the thermal near-field, the topics of thermal rectification and persistent heat flow have been more extensively explored ~\cite{zhuPersistentDirectionalCurrent2016, fanNonreciprocalRadiativeHeat2020}. Here, we extend results on monochromatic thermal currents to the far-field regime for the previously discussed reciprocal and nonreciprocal configurations, both at and out of equilibrium. Owing to translational invariance and the symmetric emissivity profile along the vertical axis, the thermal current possesses no axial component. As a result, the monochromatic heat flux vector field $\mathbf{v}$ is entirely confined to the transverse xy-plane.

\begin{figure}[t]
    \centering    \includegraphics[width=\linewidth]{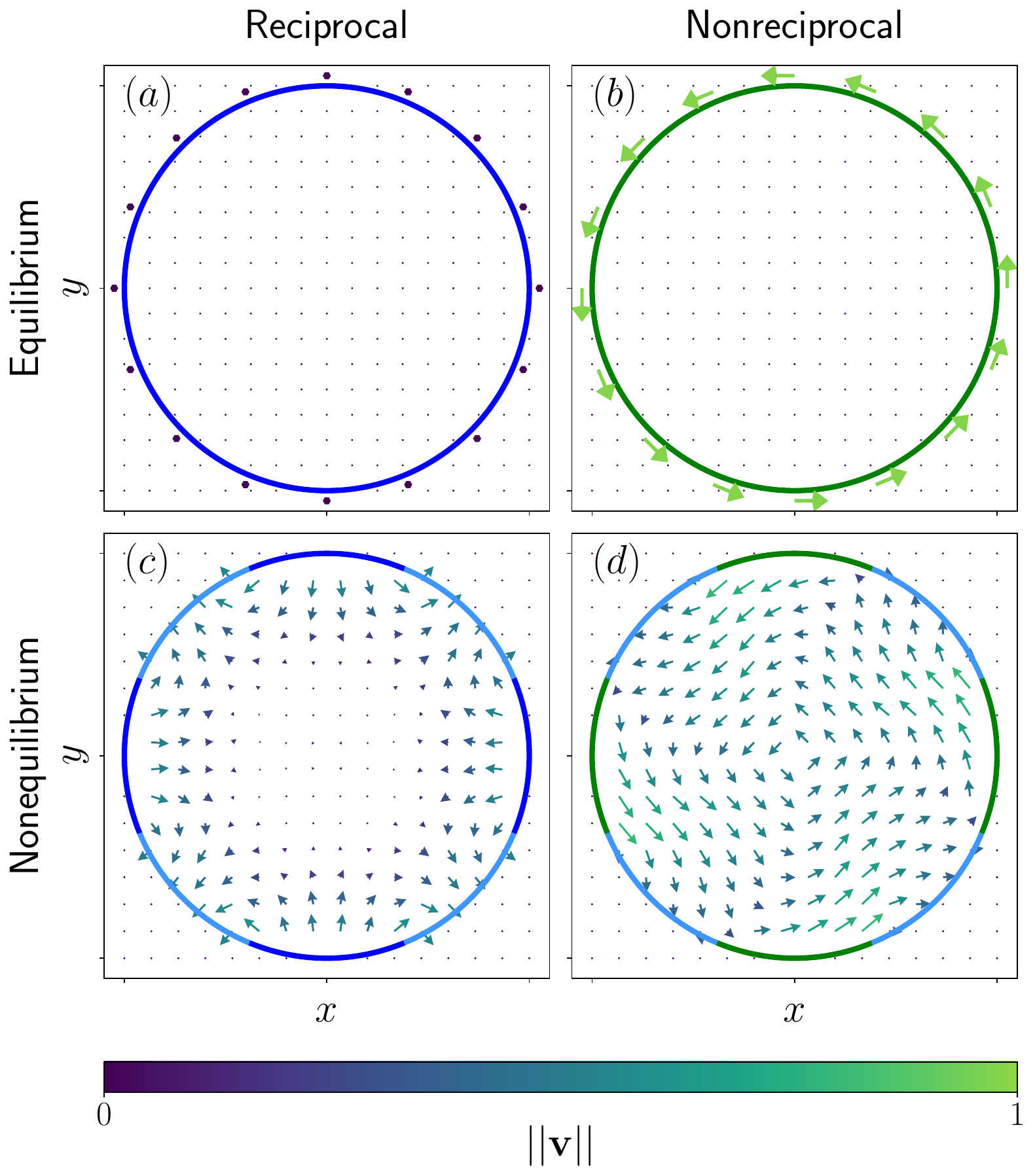}
    \caption{Monochromatic heat flux vector fields in an eight-element cylindrical configuration under different thermal and material conditions. (a) At thermal equilibrium, with all elements composed of reciprocal blackbody material, no net thermal currents are observed, either in the interior or at the boundary. (b) At thermal equilibrium, with a nonreciprocal \textit{on-off} elements (green), the internal thermal currents vanish, but a nonzero boundary inflow monochromatic heat appears, reflecting a directional imbalance in surface absorption and emission due to the asymmetric emissivity profile. (c) Out of equilibrium, with reciprocal blackbody elements at even indices (light blue, cold) and odd indices (blue, hot), the heat-flux vectors point symmetrically from hot to cold elements, yielding no preferred rotational direction. (d) In the same nonequilibrium configuration, but with nonreciprocal elements (green, hot), the heat flux vectors form a counterclockwise rotational pattern, driven by an asymmetric emissivity that favors emission into the right hemisphere. }
    \label{fig:heat_currents}
\end{figure}

The results are shown in Fig.~\ref{fig:heat_currents}. Panels (a) and (b) correspond to equilibrium conditions. Panel (a) simulates heat flux corresponding to the fully reciprocal blackbody case of Fig.~\ref{fig: Transmission coefficient graphs}(a). As expected, the simulation yields vanishing internal thermal currents and zero net boundary flux, fully consistent with thermodynamic equilibrium.

Panel (b) shows the \textit{on–off} nonreciprocal case at equilibrium, corresponding to Fig.~\ref{fig: Transmission coefficient graphs}(b). Although one might expect persistent thermal currents (i.e., heat circulation at thermal equilibrium) in the cylindrical cavity, similar to near-field many-body systems~\cite{zhuPersistentDirectionalCurrent2016, zhuTheoryManybodyRadiative2018}, no such effect is observed here. In specularly reflecting media, radiation in a given direction $(\theta, \varphi)$ consists of both direct emission and reflected rays coming from the opposite direction $(\theta, \varphi + \pi)$. Owning to the constraint $\epsilon(\omega, \theta, \varphi) = \alpha(\omega ,\theta, \varphi + \pi)$, strong emission in one direction necessarily implies strong absorption (i.e., weak reflection) from the opposite direction, and vice-versa. As a result, the total outgoing radiation defined as the sum of the emitted and reflected components, is independent of material properties. Hence, the magnitude of the monochromatic heat flux within the cavity remains isotropic, causing internal thermal currents to vanish, due to Eq.~\eqref{eq:flux-interior}. This cancellation applies for any arrangement of reciprocal and nonreciprocal materials and directly reflects the directional balance imposed by the generalized Kirchhoff’s law in Eq.~\eqref{eq:specular_emission}, which holds regardless of the material's reciprocal or nonreciprocal character. Nevertheless, a preferential direction of emission persists, as underlined by the presence of a nonzero tangential component at the boundary, even at thermal equilibrium.

Finally, we proceed to analyze the internal thermal currents in a cylindrical configuration under nonequilibrium conditions, where the even-numbered elements (light blue, panels (c) and (d)) are modeled as reciprocal blackbody materials maintained at a significantly lower temperature than the remaining elements. As such, their thermal emission is neglected. By contrast, the odd-numbered elements are assigned either reciprocal or nonreciprocal emission properties. When the hot elements are reciprocal (dark blue, panel (c)) the monochromatic heat flux vector field exhibits directional symmetry: heat flows from each hot element toward the adjacent cold elements in both directions, resulting in a balanced exchange. In contrast, when the hot elements are composed of nonreciprocal materials (green, panel (d)), the heat flux vector field organizes into a counterclockwise circulation. This rotational behavior originates from the asymmetric enhancement of emissivity into the right hemisphere, which originates from the nonreciprocal \textit{on-off} emissivity profile selected earlier.

Our results are fully consistent with the fundamental laws of thermodynamics. At thermal equilibrium, the Poynting vector is divergence-free, and the heat flux remains tangential to the boundary, ensuring the absence of net energy exchange. By contrast, under nonequilibrium conditions, energy flows from the surface of the hotter elements to the colder ones, thereby the heat flux is no longer tangential, resulting in a net energy transfer, as illustrated in Fig.~\ref{fig:heat_currents}.

\section{Conclusion}\label{Sec: Conclusions}

We have investigated radiative heat transfer in a cylindrical cavity composed of both reciprocal and nonreciprocal materials, developing a numerical framework that discretizes the cavity into vertical elements and incorporates a ray-tracing algorithm. Our results demonstrate that nonreciprocity can induce maximal heat rectification coefficients between the cylinder’s elements, even when only a subset of them breaks reciprocity while the rest remain reciprocal. At thermal equilibrium, internal thermal currents vanish in both reciprocal and nonreciprocal cases, due to the balance imposed by specular reflection. Out of equilibrium, however, nonreciprocity gives rise to directional, rotating heat fluxes within the cavity. These results advance the understanding of nonreciprocal heat transfer in the context of equilibrium but also \textit{out} of equilibrium conditions that further contribute to thermal radiation control. Extensions of this work will include the study of more complex geometries that emulate realistic configurations where the observed phenomena may be enhanced under realistic nonreciprocal conditions, such as concentric multi-cavity systems, as well as the interplay between near-field and far-field interactions by varying the radius ratio in the configuration considered. In addition, incorporating thermal conduction and time-dependent effects to capture dynamic heat transport will be crucial for establishing a realistic parameter space of materials properties for practical applications~\cite{JuNonreciprocalHeatCirculationMetadevices2024, zhuTheoryManybodyRadiative2018}.

\section*{Data availability}
The numerical simulations, codes, data, and figures used throughout the article are publicly available in G.M.S's \href{https://github.com/GuillemMasdemont/Nonreciprocal-Cavity-Paper.git}{GitHub repository}.

\begin{acknowledgments}
The authors declare no competing financial interest. G.T.P. acknowledges financial support from the la Caixa Foundation (ID No. 00010434).  This work was partially funded by CEX2024-001490-S (MICIU/AEI/10.13039/501100011033), the Spanish MICINN (PID2021-125441OA-I00,
PID2020-112625GB-I00, and CEX2019-000910-S), the European Union (Fellowship No. LCF/BQ/PI21/11830019 under Marie Skłodowska-Curie Grant Agreement No. 847648 and CATHERINA, 101168064), the Generalitat de Catalunya (2021 SGR 01443) through the CERCA program, Fundació Cellex, and Fundació Mir-Puig. Views and opinions expressed are those of the authors only and do not necessarily reflect those of the European Union or European Defence Agency.


\end{acknowledgments}

\section*{Appendices}
\appendix

\section{Computation of the transmission coefficient and numerical implementation}\label{Sec: computation S appendix}

We detail the computation of the monochromatic transmission coefficient $S_{j \to i}(\omega)$ (see Eq.~\eqref{eq:Sij}) and the monochromatic heat flux vector field $\mathbf{v}(\mathbf{p}, \omega)$ (see Eqs.~\eqref{eq:flux-interior} and \eqref{eq:flux-boundary}). In the first case, the cylindrical boundary is discretized into vertical surface elements $\Omega_i$ following Eq.~\eqref{eq cylinder partition}. Each element is further subdivided into a set of mesh cells along its width, forming a one-dimensional mesh at a fixed height, owing to the z-invariance of the cylinder. Each cell is approximated as a locally flat differential surface acting as an independent emitter. Radiation from each cell is sampled by launching a large ensemble of rays distributed over the inward hemisphere. The intensity of each ray is weighted by the prescribed directional emissivity profile. The distribution of directions is obtained through a regular local ($dk_{u}$, $dk_z$) grid, under the constraint $k_u^2 + k_z^2 \leq \lVert \mathbf{k} \rVert^2
$, where $u$ is the direction orthogonal to $z$ in the tangential plane. Lambert’s cosine law can be incorporated through the standard change of variables from momentum space to angular coordinates.


The propagation of rays is performed with a dedicated ray-tracing algorithm adapted to the cylindrical geometry. The algorithm accounts for successive reflections and transmissions at material interfaces, while consistently enforcing the decomposition introduced in Eq.~\eqref{eq: S decomposition}. To ensure numerical stability, the trajectory of a given ray is truncated once its residual intensity falls below $10^{-5}$.

In practice, simulations employ a spatial mesh of approximately $50$ cells per surface element $\Omega_i$. From each cell, $10^4$ rays are launched with uniformly sampled directions. This resolution provides an accurate approximation of the underlying angular integrals while maintaining computational feasibility. A sufficiently large number of rays is essential to guarantee convergence, since approximating the integral by a Riemann sum introduces an error that typically decays as $\mathcal{O}(n^{-1})$, where $n$ denotes the number of discretization points~\cite{burdenNumericalAnalysis2011}.

A distinct procedure is employed to compute $\mathbf{v}(\mathbf{p}, \omega)$, requiring a modification of the ray-tracing algorithm. Instead of launching rays outward from each point $\mathbf{p}$, we evaluate the contributions coming from uniformly sampled directions. In such case, the analytical expression for the monochromatic heat flux in a given direction $\hat{\mathbf{s}}$ at a point $\mathbf{p} \in \mathcal{C}$ (see Eq.~\eqref{eq:flux-interior}) can be written as
\begin{equation}\label{eq: Transmission coefficient analytically}
S(\mathbf{p}, \hat{\mathbf{s}}, \omega) 
= 
\sum_{i=1}^{\infty} 
\left( \prod_{j=1}^{i-1} R(\mathbf{r}_j, \theta, \varphi + \pi) \right) 
\epsilon(\mathbf{r}_{i}, \theta, \varphi).
\end{equation}
Here, the infinite series accounts for successive contributions from radiation emitted at $\mathbf{r}_i$ that, after undergoing $i-1$ reflections at intermediate points $\mathbf{r}_j$ ($j \in \{1,\dots,i-1\}$), arrives at $\mathbf{p}$ in direction $\hat{\mathbf{s}}$. The propagation direction $(\theta,\varphi)$ is uniquely determined by the pair $(\mathbf{p},\hat{\mathbf{s}})$ and is consistently preserved across emission and reflection events. Specifically, a ray absorbed at $\mathbf{p}$ from direction $\hat{\mathbf{s}}$ must have been emitted at $\mathbf{r}_i$ in a given direction $(\theta,\varphi)$, and reflected at intermediate points with directions $(\theta,\varphi+\pi)$. Absorbed and reflected intensities are evaluated in accordance with the constraint imposed by Eq.~\eqref{eq:specular_emission}. For the monochromatic boundary heat-flux vector field in Eq.~\eqref{eq:flux-boundary}, we include the absorptivity factor in Eq.~\eqref{eq: Transmission coefficient analytically}.
\section{Unrealistic thermal currents arising from zero-emissivity profiles}\label{sec: Spurious results}

\begin{figure}
    \centering
    \includegraphics[width=0.8\linewidth]{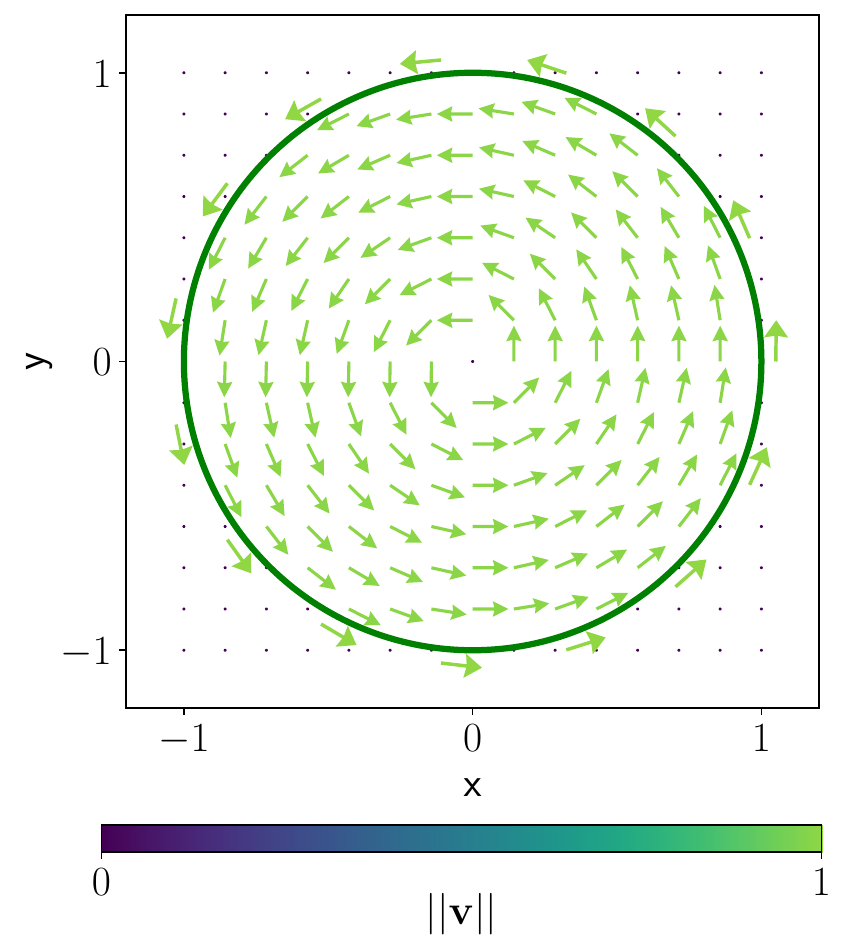}
    \caption{Monochromatic heat flux vector field in an eight-element cylindrical configuration using an abrupt, nonphysical emissivity profile defined by $\epsilon(\beta) = 1$ for $\beta \geq 0$ and $\epsilon(\beta) = 0$ for $\beta < 0$. This artificial discontinuity leads to unbalanced emission and absorption: rays are strongly emitted in the right hemisphere but cannot be absorbed from the same directions. As a result, the simulation produces unrealistic internal thermal currents that violate detailed balance.}
    \label{fig: spurious results}
\end{figure}

We note that unrealistic results arise when the emissivity is unitary in one hemisphere, and zero in the other. This phenomenon is illustrated in Fig.~\ref{fig: spurious results}. In our ray-tracing model, a configuration where the surface emits only into the right hemisphere and not into the left is unrealistic: it eliminates any radiation that would propagate clockwise, causing the appearance of internal thermal currents that are not observed in other cases. Indeed, when the emissivity is low but nonzero in the left hemisphere (as discussed in Sec.~\ref{Sec: Numerical simulations}), the specular relation $\epsilon(\omega,\theta,\varphi) = \alpha(\omega,\theta,\varphi+\pi)$ ensures a small but finite absorptivity for radiation incident from the right hemisphere. Because rays emitted into the left hemisphere inevitably reach the right hemisphere, they experience strong reflection, such that even weak left-hemisphere emission is efficiently recirculated within the cavity. This mechanism restores closed radiative exchange channels in all directions, thereby suppressing net thermal currents and producing the results displayed in Fig.~\ref{fig:heat_currents}.

\section{Numerical results for a realistic Weyl semimetal}\label{sec: weyl semimetal}

We employ a Weyl semimetal in the Voigt configuration~\cite{zhaoAxionFieldEnabledNonreciprocalThermal2020, wangMaximalViolationKirchhoffs2023} to simulate realistic radiative exchange profiles. This material is nonmagnetic (\(\mu = \mu_0\)) and characterized by a complex, anisotropic dielectric permittivity tensor. In this work we use~\cite{wangMaximalViolationKirchhoffs2023}
\begin{equation}\label{eq:Weyl_tensor}
    \epsilon =
    \begin{pmatrix}
        9 + 0.3i & 0 & 9i \\
        0 & 9 + 0.3i & 0 \\
        -9i & 0 & 9 + 0.3i
    \end{pmatrix}.
\end{equation}
Weyl semimetals in this configuration break Lorentz reciprocity without the application of an external magnetic field. The reflection coefficients $r_s$ and $r_p$ at the air - Weyl semimetal interface are evaluated numerically using the \texttt{GeneralizedTransferMatrixMethod.jl}. The mean absorption profile is directly obtained from the mean reflection coefficient, while the mean emissivity profile is computed via Eq.~\eqref{eq:specular_emission}, as the two are related in specular media~\cite{YangPolarimetricanalysisthermalemissionbothreciprocalnonreciprocalmaterialsusingfluctuationelectrodynamics2022}.

Qualitatively, the behavior of Weyl semimetals resembles that of the idealized \textit{on-off} emitter, though with significantly reduced emission asymmetry. For instance, the three element configuration shown in Fig.~\ref{fig:3 element configuration - WS} yield a heat rectification coefficient of approximately $H_{ij} = 0.028$. Similarly, in a full cylindrical configuration analogous to Fig.~\ref{fig: Transmission coefficient graphs}(b), the resulting heat rectification coefficient $H_{ij}$ is approximately an order of magnitude smaller and yields a nonreciprocity factor of $\zeta = 0.061$ (9.0 \% of the idealized on–off emitter.). Note that of course, these values are subject to change for Weyl semimetals with different permittivities.

\bibliography{MyLibrary}

\end{document}